# Simple sampling clock synchronisation scheme for reduced-guard-interval coherent optical OFDM systems

O. Omomukuyo, D. Chang, O. A. Dobre, R. Venkatesan and T. M. N. Ngatched

A simple data-aided scheme for sampling clock synchronisation in reduced-guard-interval coherent optical orthogonal frequency division multiplexing (RGI-CO-OFDM) systems is proposed. In the proposed scheme, the sampling clock offset (SCO) is estimated by using the training symbols reserved for channel estimation, thus avoiding extra training overhead. The SCO is then compensated by resampling, using a time-domain interpolation filter. The feasibility of the proposed scheme is demonstrated by means of numerical simulations in a 32-Gbaud 16-QAM dual-polarisation RGI-CO-OFDM system.

*Introduction:* Orthogonal frequency division multiplexing (OFDM) has been actively investigated in recent years for optical communications because of its advantages including high spectral efficiency and robustness to both chromatic dispersion (CD) and polarisation mode dispersion [1]. Reduced-guard-interval coherent optical OFDM (RGI-CO-OFDM) [2], which combines coherent optical detection and OFDM with a short cyclic prefix (CP) length, is regarded as a promising candidate for future high-speed, long-haul optical transmission systems.

It is well-known that OFDM systems are sensitive to synchronisation errors. In contrast to frame and frequency synchronisation, for which several schemes have been proposed for coherent optical OFDM (CO-OFDM), see e.g. [3], sampling clock synchronisation has not received as much research attention. Sampling clock synchronisation involves estimating and compensating for the sampling clock offset (SCO) which is caused by the mismatch in the frequencies of the sampling clocks of the digital-to-analogue converter (DAC) at the transmitter (Tx) and the analogue-to-digital converter (ADC) at the receiver (Rx). The SCO results in amplitude reduction, phase rotation and inter-carrier interference (ICI) [4]. Typically, in most existing CO-OFDM work, SCO is avoided by either manually adjusting the arbitrary waveform generator and the scope, or by using a common external reference clock to drive the DAC and ADC. It is obvious that these approaches will not be feasible in real optical transmission systems which have a standard SCO of 200 parts per million (ppm) and physically-separated Txs and Rxs [5]. Another approach for sampling clock synchronisation is to transmit a dedicated clock signal to the Rx [6]. This approach is disadvantaged by the requirement for additional circuits in the Tx and Rx. A digital signal processing (DSP) method has been proposed by Yi and Qiu in [7]. However, in that method, an SCO estimation error between 5% and 300% is observed [7].

In this letter, we propose a novel simple data-aided scheme to achieve sampling clock synchronisation, where the SCO is estimated by using the channel training symbols (TSs). The SCO estimate is then fed back to a time-domain interpolation filter module where the SCO is corrected by resampling. The proposed scheme does not require any additional hardware or training overhead and can compensate for up to the standard SCO of 200 ppm with low estimation error and negligible optical signal-to-noise ratio (OSNR) penalty.

*System model and operating principle:* Fig. 1 shows the setup of the dual-polarisation RGI-CO-OFDM system used to investigate the performance of the proposed scheme. In the case that the sampling clocks of the DAC and the ADC are not synchronised, the received samples in the $l$th OFDM symbol can be expressed as:

$$r(n) = r(nT_t + n\gamma T_t), \qquad (1)$$

where $n$ is the received sample index, $T_t$ is the sampling period of the DAC clock, $\gamma = (T_r - T_t)/T_t$ is the relative SCO, and $T_r$ is the sampling period of the ADC clock. The relationship between $n$ and $l$ can be expressed as follows:

$$n = lN_s + N_{CP}, lN_s + N_{CP} + 1, \cdots, lN_s + N_{CP} + N - 1,$$

$$l = 0, 1, \cdots, L-1, \qquad (2)$$

where $N$ is the inverse fast Fourier transform (IFFT) size, $N_{CP}$ is the CP duration, $N_s = N + N_{CP}$ is the OFDM symbol length, and $L$ is the total number of OFDM symbols in each frame. From (1), it is clear that the SCO causes each received sample in the $l$th OFDM symbol to be shifted from its original position by $n\gamma T_t$. After demodulation of the subcarriers via a fast Fourier transform (FFT), this shift results in a phase rotation of $2\pi k \Delta f_N n\gamma T_t$ of the $k$th OFDM subcarrier in the $l$th OFDM symbol, where $\Delta f_N$ is the OFDM subcarrier frequency spacing. For RGI-CO-OFDM systems, and for large-enough values of $l$, $n \cong lN_s$. Consequently, this phase rotation, $\varphi_{k,l}$, can be approximated as:

$$\varphi_{k,l} \cong s_l k$$
$$k = \frac{-N_{sc}}{2} + 1, \ \frac{-N_{sc}}{2} + 2, \cdots, \frac{N_{sc}}{2}, \qquad (3)$$

where $s_l = (2\pi l N_s \gamma)/N$ is the phase shift slope of the $l$th OFDM symbol, and $N_{sc} (\leq N)$ is the number of OFDM subcarriers. It can be seen from (3) that the phase rotation caused by the SCO increases in a linear fashion with both the OFDM subcarrier and symbol indices.

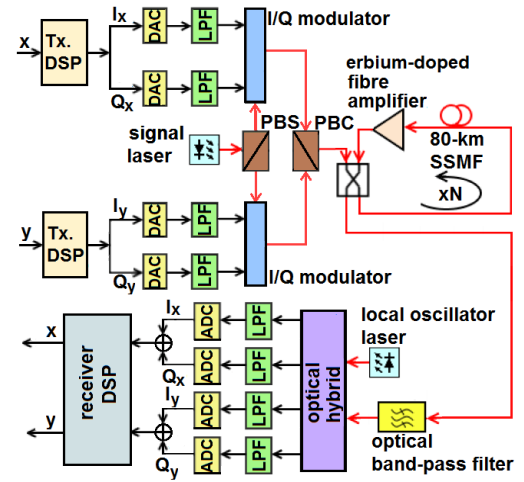

**Fig. 1** *Setup of the dual-polarisation RGI-CO-OFDM system*
LPF: low-pass filter. PBS: polarisation beam splitter. PBC: polarisation beam coupler. SSMF: standard single-mode fibre

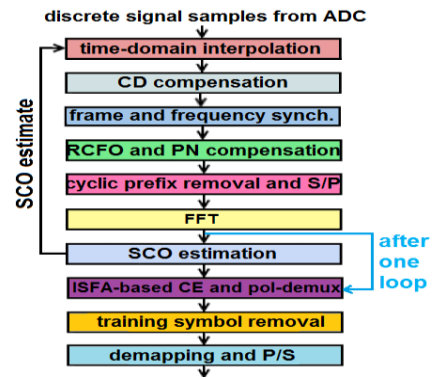

**Fig. 2** *Receiver DSP structure*
Synch: synchronisation. RCFO: residual carrier frequency offset. PN: phase noise. S/P: serial-to-parallel conversion. ISFA: intra-symbol frequency-domain averaging. CE: channel estimation. Pol-demux: polarisation demultiplexing. P/S: parallel-to-serial conversion

Fig. 2 shows the Rx DSP structure of the RGI-CO-OFDM system. After CD compensation, frame and frequency synchronisation, as well as residual CFO and phase noise compensation have all been carried out (using appropriate existing algorithms), the channel TS is used to estimate the SCO. After carrying out the FFT, the phase rotation, $\varphi_{k,TS}$, caused by the SCO on the TS can be obtained as:

$$\varphi_{k,TS} = arg\left\{\frac{r_{k,TS}}{x_{k,TS}}\right\}, \tag{4}$$

where $r_{k,TS}$ and $x_{k,TS}$ are the received and transmitted baseband symbols on the $k$th OFDM subcarrier in the TS, respectively. A least-squares curve fit of $\varphi_{k,TS}$ can then be obtained, from which the slope is computed, and then, $\gamma$ is estimated using (3). The estimate of $\gamma$ is fed back to the time-domain interpolation filter module and then, the SCO is compensated by resampling. The SCO estimation module is bypassed after one loop of feedback processing to correct the SCO.

*Simulation results:* In Fig. 1, the optical model is built using VPI TransmissionMaker, while all DSP is performed in MATLAB. At the Tx, for each polarisation branch, a data stream consisting of a deBruijn sequence of length $2^{19}$ is mapped onto 412 OFDM data subcarriers with 16-QAM encoding. These data subcarriers are converted to the time-domain using a 512-point IFFT with a CP length of 46 samples. A pair of correlated dual-polarisation TSs is used for channel estimation (CE) and SCO estimation, and intra-symbol frequency-domain averaging is utilised to improve the CE accuracy. The signal laser has a linewidth of 100 kHz and centre emission wavelength of 1550 nm, while the DACs have sampling rates of 40 GSa/s. The optical signal is launched into a recirculating loop consisting of 80-km standard single-mode fibre and a 16-dB gain erbium-doped fibre amplifier. The amplified spontaneous emission noise is suppressed by a 0.8-nm bandwidth optical band-pass filter. At the Rx, a laser with a linewidth of 100 kHz and a carrier frequency offset of 5 GHz is used as the local oscillator. The coherently-detected signal is sampled by 8-bit resolution ADCs with sampling rates set between 39.992 GSa/s and 40.008 GSa/s to emulate SCOs between -200 and 200 ppm.

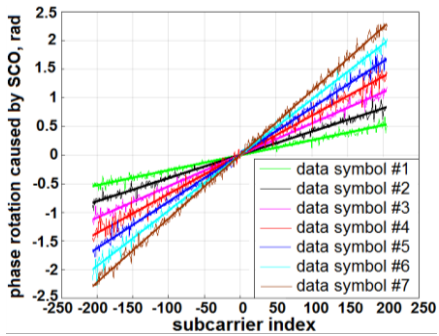

**Fig. 3** *Phase rotation caused by SCO versus subcarrier index*

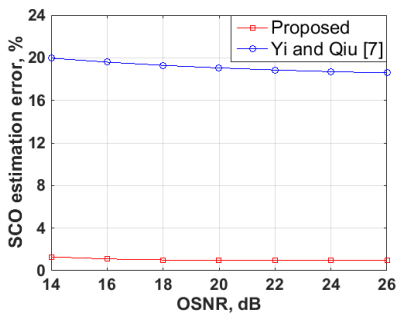

**Fig. 4** *SCO estimation error versus OSNR for an SCO of 200 ppm*

Fig. 3 shows the phase rotation caused by an SCO of 200 ppm for the first 7 data symbols for the x-polarisation branch. The least-squares fits of the actual phase rotations are indicated by the thicker lines in the figure. It can be seen that in agreement with (3), the phase rotation of the received signal increases linearly with both the symbol and subcarrier indices. Fig. 4 shows that the proposed scheme has a smaller SCO estimation error than the Yi and Qiu's method [7] over a wide range of OSNRs. Fig 5*a* shows the bit error rate (BER) without forward error correction against the OSNR for the cases when the SCO varies from -200 ppm to 200 ppm, including the case when the SCO is 0 ppm.

It can be seen that for all SCO values, there is only a negligible OSNR penalty obtained when the proposed scheme is utilised. Fig. 5*b* shows that when no SCO compensation is in place, an increase in the SCO results in significant BER degradation because the SCO-induced ICI gets worse. In contrast, when the proposed scheme is used, the BER remains basically constant for all SCO values.

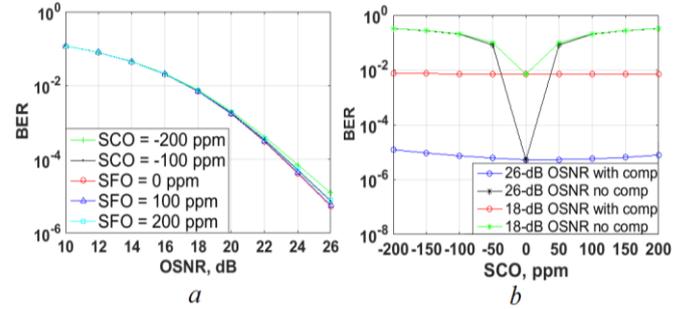

**Fig. 5** *BER curves for the x-polarisation branch for 800-km standard single-mode fibre transmission*
*a* BER vs. OSNR for various SCOs
*b* BER vs. SCO for 18-dB and 26-dB OSNR

*Conclusion:* A novel scheme for sampling clock synchronisation has been proposed, and its performance numerically demonstrated in a 32-Gbaud 16-QAM dual-polarisation RGI-CO-OFDM system. The proposed scheme can compensate for up to the standard SCO of 200 ppm without incurring significant OSNR penalty or requiring any additional training overhead or hardware, making it attractive for potential implementation in real coherent optical systems.